\documentclass[aps,twocolumn,groupedaddress,showpacs]{revtex4}
\usepackage{graphicx}
\newcommand\beq{\begin{eqnarray}}
\newcommand\eeq{\end{eqnarray}}
\newcommand\bq{\begin{equation}}
\newcommand\eq{\end{equation}}

\begin{document}
\title{The polarized electron target as a new solar-neutrino detector}

\author{M. Misiaszek}
\affiliation{ M. Smoluchowski
Institute of Physics, Jagiellonian University, ul. Reymonta 4,\\
PL-30-059 Krak\'ow, Poland }
\author{S. Ciechanowicz}
\author{W. Sobk\'ow}
\affiliation{Institute of
Theoretical Physics, University of Wroc\l{}aw, Pl. M. Born 9,
\\PL-50-204~Wroc\l{}aw, Poland}

\date{\today}

\begin{abstract}
In this paper, we analyze the scattering of solar neutrinos on
the polarized electron target, and predict how the effect of parity violation
in weak interactions may help to distinguish neutrino signal from detector background.
We indicate that the knowledge of the Sun motion across the sky 
is sufficient to predict the day/night asymmetry in the $(\nu_ee^-)$ scattering on
the polarized electron target. To make this detection feasible, the polarized electron target for solar neutrinos needs to be build from magnetic materials, e.g. from ferromagnetic iron foils, paramagnetic scintillator crystals or scintillating ferrofluids.

\end{abstract}

\maketitle

\section{\label{Sec1}Introduction}
The solar neutrinos have been detected by several underground detectors \cite{davies,gno,sage,kamiokande,sk,sno}.
These are milestones of modern astrophysics and particle physics, which gave us a unique opportunity to look inside the Sun. The water Cherenkov detectors are sensitive to
the direction of the outgoing lepton, thus giving information on the directional dependence
of neutrino events. The methods used for low-energy neutrino detection are not 
sufficiently selective for identification of the absolute neutrino flux.
The main problem is related with very large background for rare neutrino events \cite{wojcik1,wojcik2,wojcik3}.
We indicate that the isotropic background rate can be distinguished from the solar neutrino interactions, because left-handed solar neutrinos are mainly interacting with left-handed electrons. 
If a right-handed electrons are exposed to solar neutrino flux, the neutrino event rate decreases
while the detector background stays the same.  We can ''switch off'' the Sun to measure the 
background level.
\par To determine the flux of low energy neutrinos, we need to construct a low-threshold, real-time, solar-neutrino detector in which polarization of the electron targets can be controlled. 
The scintillators have a very good response to low-energy electrons. In ferro- and para-magnetic materials an electrons are polarized when magnetic field is applied. The detector should be constructed from magnetic material and scintillating media.

\section{\label{Sec2}Laboratory differential cross section}
In this section, we consider the advantages of the $(\nu_e e^-)$ scattering,
when the incoming solar neutrino beam consists only of the L-handed and
longitudinally polarized neutrinos. We assume
that these neutrinos are detected in the standard $(V-A)_L$  weak
interactions with the polarized electron target (PET) and the recoil electron  
energy spectrum is measured. We consider the case when the outgoing
electron momentum direction is not observed, because of low electron recoil 
energy deposited at a short distance.
The formula for the laboratory differential
cross section \cite{Yang,Rashba,Minkowski,nasza} is presented 
after integration over the azimuthal
angle $\phi_{e'}$ of the recoil electron momentum (see Fig.1): 
\begin{widetext}
\beq  \label{cross3VA} \left(\frac{d \sigma}{d y }\right)_{(V, A)}
&=& \frac{E_{\nu}m_{e}}{2\pi}\frac{G_{F}^{2}}{2}(1-\mbox{\boldmath
$\hat{\eta}_{\nu}$}\cdot\hat{\bf q})\Bigg\{\left(c_{V}^{L} +
c_{A}^{L}\right)^{2}(1 + \mbox{\boldmath
$\hat{\eta}_{e}$}\cdot{\bf\hat{q}})  + \left(c_{V}^{L} -
c_{A}^{L}\right)^{2}\left[1 - (\mbox{\boldmath
$\hat{\eta}_{e}$}\cdot {\bf \hat{q}})\left(1-
\frac{m_{e}}{E_{\nu}}\frac{y}{(1-y)}\right)\right](1-y)^{2}
\nonumber\\
& & \mbox{} - \left[\left(c_{V}^{L}\right)^{2} -
\left(c_{A}^{L}\right)^{2}\right] (1 +
\mbox{\boldmath$\hat{\eta}_{e}$}\cdot{\bf\hat{q}})\frac{m_{e}}{E_{\nu}}y\Bigg\},
\eeq
\end{widetext}
where $\mbox{\boldmath $\hat{\eta}_{\nu}$}\cdot\hat{\bf q} = -1 $
is the longitudinal polarization of the incoming L-handed
solar neutrino, ${\bf q}$ - the incoming neutrino momentum, $ \mbox{\boldmath
$\hat{\eta}_{e}$}$ - the unit 3-vector of the initial electron
polarization in its rest frame. The measurement of the projection of the electron polarization vector parallel
to neutrino direction  $\mbox{\boldmath$\hat{\eta}_{e}$}\cdot{\bf\hat{q}}$ is only possible when the electron target polarization is known. The
polarization vector for electrons is parallel to the applied magnetic
field. The variable $y $ is the ratio of the kinetic energy
of the recoil electron $T_{e} $ to the incoming neutrino energy
$E_{\nu} $:
\beq
y\equiv\frac{T_{e}}{E_{\nu}}=\frac{m_{e}}{E_{\nu}}\frac{2cos^{2}\theta_{e'}}
{(1+\frac{m_{e}}{E_{\nu}})^{2}-cos^{2}\theta_{e'}}. \eeq
It varies from $0 $ to $2/(2+m_e/E_\nu) $. $\theta_{e'}$ - the
polar angle between the direction of the outgoing electron
momentum  $\hat{\bf p}_{e'}$ and the direction  of the incoming
neutrino momentum $\hat{\bf q}$ (recoil electron scattering
angle), $m_{e}$ - the electron mass. The experimental values
of the standard coupling constants are: $c_{V}^{L}=1-0.040\pm
0.015 $, $c_{A}^{L}=1-0.507\pm 0.014 $ \cite{Data}, when the charged
current weak interaction is included.

\begin{figure}
\begin{center}
\includegraphics[scale=.7]{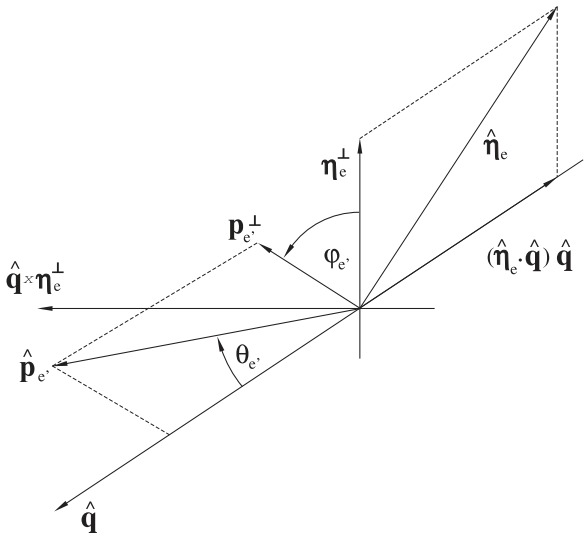}
\end{center}
\caption{Figure shows the reaction plane for the $(\nu_e e^-)$
scattering on the PET, $\mbox{\boldmath $\hat{\eta}_{e}$}$ - the unit
3-vector of the electron target polarization in its rest frame.} \label{pet}
\end{figure}

The electron recoil spectrum depends on the initial electron polarization. For $^7Be$ 
solar neutrino scattering on the PET with energy $E_{\nu}=0.862 \, \mbox{MeV}$, we present
in Fig.2 three interesting cases: $\mbox{\boldmath$\hat{\eta}_{e}$}\cdot{\bf\hat{q}} = -1,0,1$. 
When the polarization vector points to the Sun ($\mbox{\boldmath$\hat{\eta}_{e}$}\cdot{\bf\hat{q}} = -1$) the total cross section
decreases to less than $5 \%$ of its value in the unpolarized case ($\mbox{\boldmath$\hat{\eta}_{e}$}\cdot{\bf\hat{q}} = 0$).
If the polarization vector is parallel to the solar neutrino momentum vector and points to the same direction ($\mbox{\boldmath$\hat{\eta}_{e}$}\cdot{\bf\hat{q}} = 1$) the total cross section is $200 \%$ of the
unpolarized case ($40 \times$ times larger then the $\mbox{\boldmath$\hat{\eta}_{e}$}\cdot{\bf\hat{q}} = -1$ case).
In the total cross section calculation we integrate Eq. (\ref{cross3VA}) over electron recoil energies from $y = 0.2$ (e.g.
the detector energy threshold) to $y=0.77$ (the kinematic maximum).

\par It can be noticed that if the initial electron polarization vector $ \mbox{\boldmath
$\hat{\eta}_{e}$}$ is fixed in the laboratory frame the value of projection $\mbox{\boldmath$\hat{\eta}_{e}$}\cdot{\bf\hat{q}}$  
is varying in time due to diurnal motion of the Sun across the sky (an effect of the Earth's
rotation). Indeed, if the $ \mbox{\boldmath
$\hat{\eta}_{e}$}$ vector points to the zenith the value of $\mbox{\boldmath$\hat{\eta}_{e}$}\cdot{\bf\hat{q}}$ is equal to $\cos(Alt + \frac{\pi}{2})$, where $Alt$ is the angle (Altitude) between the horizon and the Sun at any given instant. 
Hence, it is easy to calculate the day/night asymmetry in the $(\nu_ee^-)$ scattering on
the PET using ephemeris data. It would be clear signature of solar neutrinos. This asymmetry does vanish,
 if electrons are unpolarized. 
From the point of view of our method, the rate of the scattering of solar neutrinos on unpolarized electrons 
should be treated as an isotropic background. Only polarized electrons are detection targets.

\begin{figure}
\begin{center}
\includegraphics[scale=.33]{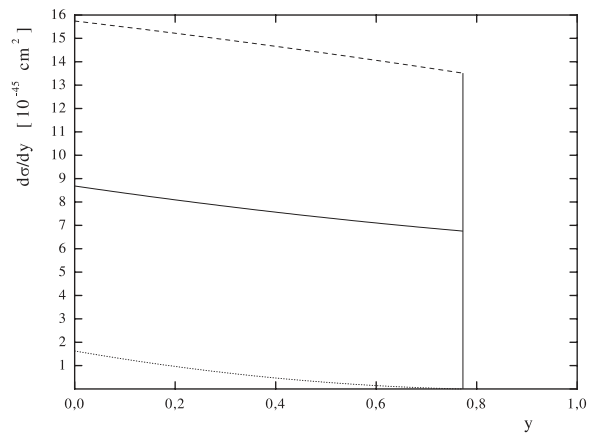}
\end{center}
\caption{Plot of the $\frac{d \sigma}{d y}$ as a
function of $y$ for the $(\nu_{e}e^{-})$ scattering of L-handed $^7Be$ solar neutrinos
on the PET, $E_{\nu}=0.862\, \mbox{MeV}$; 
a) the case of $\mbox{\boldmath$\hat{\eta}_{e}$}\cdot{\bf\hat{q}} = 0$ (solid line), 
b) $\mbox{\boldmath$\hat{\eta}_{e}$}\cdot{\bf\hat{q}} = 1$ (dashed line),
c) $\mbox{\boldmath$\hat{\eta}_{e}$}\cdot{\bf\hat{q}} = -1$ (dotted line).}
 \label{FIG2}
\end{figure}

\section{\label{PET}The feasibility of developing \newline the PET}

The polarized target electrons are produced in para- and ferro-magnetic materials
that are magnetized with using external magnetic field. 
\par In a piece of magnetized ferromagnetic material  there are lots of unpaired electrons 
all pointing the same direction. The degree of alignment of the electron spins between 
neighboring atoms is high as a result of the exchange force that tends to lock the spins 
of these electrons in a parallel direction. For example, at flux density of $\, \simeq 2 \, T$ the iron
becomes magnetically saturated, yielding a target polarization of $\, \simeq  8 \%$  \cite{JLab}.
The metallic gadolinium is another interesting ferromagnetic material with the magnetic moment per 
atom $\mu_{Gd} = 7.63 \, \mu_{B}$ and with the Curie temperature $T_C \simeq 292 \, \mbox{K}$ \cite{MetPhys}.
The detector would consist of thin ferromagnetic foils sandwiched between scintillator plates (which would
measure the energy of the event).
\par Paramagnetic materials have a small and positive susceptibility to magnetic fields.
Paramagnetic properties are due to the presence of some unpaired electrons and from the
realignment of the electron orbits caused by the external magnetic field. The $Fe^{3+}$ ion
has a $\mbox{3d}^5$ electron configuration with a Hund's rule ground state of $^6\mbox{S}_{5/2}$ resulting 
in a magnetic moment of $5 \, \mu_{B}$. The $Gd^{3+}$ ion has a $\mbox{4f}^7\mbox{5s}^2\mbox{p}^6$ electron configuration with a ground state of $^8\mbox{S}_{7/2}$ and a magnetic moment of $7 \, \mu_{B}$. 
The magnetization of single paramagnetic atoms is described by a Brillouin function.
Only at low temperatures and in strong magnetic fields the electron polarization is enough to build
the PET. For example, magnetic moment of $Gd^{3+}$ ion reaches $\simeq 6.3 \, \mu_{B}$ at $B/T \simeq 1 \, T\cdot K^{-1}$ \cite{Henry}. The paramagnetic materials are of our interest due to their potential to 
build scintillating crystals. The cerium-doped gadolinium silicate ($Gd_2SiO_5:Ce$, or GSO:Ce) is a fast and high-Z scintillator \cite{GSO1} with the light yield as large as about $20\%$ of that of $NaI(Tl)$ \cite{GSO3}.
Magnetized GSO crystal could be both the PET and high precision electromagnetic calorimeter at the same time.    
\par A ferrofluid is a stable colloidal suspension of sub-domain magnetic particles in a liquid carrier. The particles, which have an average size of about 10 nm, are coated with a stabilizing dispersing agent (surfactant) which prevents particle agglomeration even when a strong magnetic field gradient is applied to the ferrofluid. In the absence of a magnetic field, the magnetic moments of the particles are randomly distributed and the fluid has no net magnetization. When a magnetic field is applied to a ferrofluid, the magnetic moments of the particles are freely rotating and orient along the field lines almost instantly \cite{ferrotec}. Xerox ferrofluids incorporate particles of maghemite, or $gamma-Fe_2O_3$, as the magnetic species to produce some of the most optically transparent magnetic materials known for applications at ordinary room temperatures \cite{xerox}. Xerox ferrofluids may be dissolved in water Cherenkov detectors. Research work should be done to find transparent and scintillating ferrofluid. 
\par We state that the PET with the initial electron polarization is feasible. To give a numerical example, we may
consider the high resolution detector based on many thin scintillator and magnetized iron plates. 
There are $N_e \simeq 1.7 \cdot 10^{31}$ fully polarized electrons in 750 tons of metallic iron (we assume that $8\%$ of electrons are polarized at saturation). The total cross section for $^7Be$ ($E_{\nu}=0.862 \, \mbox{MeV}$ and $\phi_\nu = 0.43 \cdot 10^{10} cm^{-2} s^{-1}$ \cite{BP04}) solar neutrino scattering on the PET with electron polarization  $\mbox{\boldmath$\hat{\eta}_{e}$}\cdot{\bf\hat{q}} = 1$ and detector energy threshold $y_{th} = 0.2 $ is equal to $\sigma \simeq 8.2 \cdot 10^{-45} \, cm^{2}$. Under such conditions, the rate of $^7Be$ neutrino interactions on the PET is about $N = N_e \cdot \phi_\nu \cdot \sigma \cdot 24 \cdot 3600 \simeq 52$ events per day. 
\par

\section{Conclusions}

In the first part of the paper, we show that the diurnal asymmetry in a number of recoil 
electrons may be used to distinguish solar neutrino interactions from
detector background rate. Next, we give examples of detection techniques. 
The future neutrino detectors based on the PET
will provide the unique opportunity for low energy neutrino astronomy, if
the large magnetized sampling calorimeters are build underground.
The PET can be used as a low energy neutrino telescope.

\begin{acknowledgments}
This work was supported in part by the grant 2P03B 15522 of The
Polish Committee for Scientific Research.
\end{acknowledgments}

\end{document}